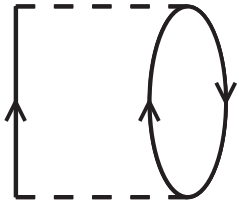 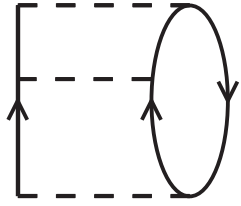 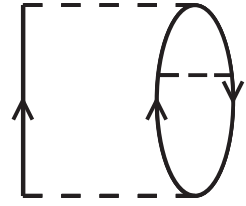

a)    b)    c)

# Long-Range Correlations and the Momentum Distribution in Nuclei

K. Amir-Azimi-Nili, H. Müther

*Institut für Theoretische Physik,*
*Universität Tübingen, D-72076 Tübingen, Germany*

L.D. Skouras

*Institute of Nuclear Physics, N.R.C.P.S. Demokritos*
*Aghia Paraskevi GR 15310, Greece*

A. Polls

*Departament d'Estructura i Constituents de la Matèria*
*Universitat de Barcelona, E-08028 Barcelona, Spain*

The influence of correlations on the momentum distribution of nucleons in nuclei is evaluated starting from a realistic nucleon-nucleon interaction. The calculations are performed directly for the finite nucleus $^{16}$O making use of the Green's function approach. The emphasis is focused on the correlations induced by the excitation modes at low energies described within a model-space of shell-model configurations including states up to the sdg shell. Our analysis demonstrates that these long-range correlations do not produce any significant enhancement of the momentum distribution at high missing momenta and low missing energies. This is in agreement with high resolution $(e,e'p)$ experiments for this nucleus. We also try to simulate the corresponding effects in large nuclei by quenching the energy-spacing between single-particle orbits. This yields a sizable enhancement of the spectral function at large momenta and small energy. Such behavior could explain the deviation of the momentum distribution from the mean field prediction, which has been observed in $(e,e'p)$ experiments on heavy nuclei like $^{208}$Pb.

# I. INTRODUCTION

A lot of effort has been made to explore the limits of the simple shell-model or independent particle model (IPM) of the nucleus. It has been argued that the strong short-range and tensor components of realistic nucleon-nucleon (NN) interactions should induce short-range correlations into the nuclear wave function. These correlations should give rise to an enhancement of the momentum distribution at high momenta as compared to the momentum distribution derived from a Hartree-Fock or mean field description of the nucleus. Therefore high-resolution $(e,e'p)$ experiments have been performed to determine the spectral function of nucleons at high momenta leading to the ground-state or states with low excitation energies in the daughter nucleus [1,2].

Microscopic calculations, which account for the effects of these short-range correlations, indeed predict components in the momentum distribution at momenta around 2-3 $fm^{-1}$, which are larger by orders of magnitude as compared to the predictions of Hartree-Fock or IPM calculations [3–6]. The momentum distribution can be obtained from integrating the spectral function for nucleon knock-out over all missing energies, i.e. all excitation energies of the remaining nuclear system. A more detailed analysis [6,7] of this spectral function showed that the high-momentum components mainly originate from the spectral function at large energies while the momentum distribution for nucleons with small energies can rather well be approximated by those derived from the IPM. Similar results are also obtained in the study of the spectral function for nuclear matter [8–11].

This analysis was essentially confirmed by the experimental data. The experimental data for the spectral function in the light nucleus $^{16}$O at low energies were in a good agreement with the prediction of an IPM [1]. On the other hand, the momentum distribution for the heavy nucleus $^{208}$Pb showed an enhancement at large momenta and small energies as compared to a Hartree-Fock or IPM prediction [2]. This enhancement is not as large as the one predicted from short-range correlations for the total momentum distribution, but the deviation from the IPM is so significant that the authors of [2] suggested that this enhancement might be due to long-range correlations, which corresponds to low-energy excitations of the many-body system. Their suggestion is supported by a simple estimate of the effect of long-range correlations on the momentum distribution by Ma and Wambach [12]. In this approach the effects of correlations are included by means of an effective mass, which depends on the position in $r$-space.

Also the fact that this enhancement of the momentum distribution is observed for a heavy nucleus but not for the light nucleus $^{16}$O supports the idea that the enhancement at small energies may originate from long-range correlations. The effects of short-range correlations should be rather independent of the nuclear system under consideration. They can possibly be derived from studies of nuclear matter assuming a local density approximation [11]. As indicated by the name these correlations are of short range and not very sensitive to the global structure of the whole nuclear system. Consequently the effects of short-range correlations should be rather similar for the nuclei $^{16}$O and $^{208}$Pb. Contrary long-range correlations could be sensitive to the whole nuclear system and give different results for different nuclei. They are related to the excitations of low energy and therefore results derived from nuclear matter, which shows a continuous single-particle spectrum, can be quite different from those in finite nuclei, for which the low-energy excitations are rather sensitive to the shell-structure.

These considerations lead us to the conclusion that the investigation of long-range correlation effects or low-energy excitation modes should be performed directly for the finite nucleus under consideration, while a study of nuclear matter may not be very reliable. Such investigations of low-energy excitations are typically performed in a model-space of the shell-model. Assuming a set of bound single-particle states, like e.g. the eigenstates of an harmonic oscillator, the basic shell-model configurations are generated and the eigenstates are obtained by diagonalizing the residual interaction between these basic configurations. Such a basis in terms of oscillator states is rather appropriate to describe the energies and the mixing of configurations for the excitation modes with low energy. A finite basis of oscillator states, however, is not at all appropriate to describe high-momentum components in the nuclear wave function because these high-momentum components will be dominated by the tail of the oscillator basis states. Therefore, as we will explain below, we have used a mixed representation of basis states, which considers a shell-model basis to describe the excitation modes, but a basis of plane-wave states to determine the momentum distribution.

The spectral function and the momentum distribution will be evaluated within the Green's function approach for the nuclear many-body theory (for a reference see e.g. the recent review articles [13,14]). After this introduction we will in section 2 briefly summarize this method and present the details of the approximations and techniques which we use in our calculation. The results will be discussed in section 3 and the main conclusions are summarized in the last section.

# II. EVALUATION OF THE SPECTRAL FUNCTION

For studies of finite systems it is convenient to introduce a partial wave decomposition of the spectral function

$$S_{lj\tau}(k, E) = \sum_n |<\Psi_n^{A-1}|a_{klj\tau}|\Psi_0^A>|^2 \; \delta(E-(E_0^A-E_n^{A-1})) \tag{1}$$

where $a_{klj\tau}$ denotes the annihilation operator for a nucleon with orbital angular momentum $l$, total angular momentum $j$, isospin $\tau$ and momentum $k$. The state $|\Psi_0^A>$ refers to the ground state of the target nucleus, while $|\Psi_n^{A-1}>$ is used to identify the various eigenstates of the hamiltonian with one particle removed from the target nucleus. Hence the hole spectral function $S(\vec{k}, E)$ gives the probability of removing a particle with momentum $\vec{k}$ from the target system of $A$ particles leaving the resulting (A-1) system with an energy $E^{A-1} = E_0 - E$, where $E_0$ is the ground state energy of the target. The momentum distribution for nucleons with corresponding quantum numbers

$$n_{lj\tau}(k) = <\Psi_0^A|a_{klj\tau}^\dagger a_{klj\tau}|\Psi_0^A> \tag{2}$$

can be rewritten by inserting a complete set of eigenstates $|\Psi_n^{A-1}>$ for the system with $A-1$ nucleons

$$n_{lj\tau}(k) = \sum_n |<\Psi_n^{A-1}|a_{klj\tau}|\Psi_0^A>|^2 \tag{3}$$

and is obtained by integrating the spectral function $S_{lj\tau}(k, E)$ over the excitation energies of the $A-1$ system. Then taking into account the degeneracy factors of each orbital one obtains the total momentum distribution

$$n(k) = \sum_{l,j,\tau} (2j+1)\; n_{lj\tau}(k). \tag{4}$$

The spectral function for the various partial waves, $S_{lj\tau}(k, E)$ can be obtained from the imaginary part of the corresponding single-particle Green's function $g_{lj}(k_1, k_2 E)$. Note that here and in the following we have dropped the isospin quantum number $\tau$, as we ignore the Coulomb interaction between the protons.

To determine the correlated single-particle Green's function one has to solve a Dyson equation, which in the case of finite systems takes an integral form

$$g_{lj}(k_1, k_2; \omega) = g_{lj}^{(HF)}(k_1, k_2; \omega) + \int dk_3 \int dk_4 g_{lj}^{(HF)}(k_1, k_3; \omega) \Delta\Sigma_{lj}(k_3, k_4; \omega) g_{lj}(k_4, k_2; \omega). \tag{5}$$

where $g^{(HF)}$ refers to a Hartree-Fock propagator and $\Delta\Sigma_{lj}$ represents contributions to the irreducible self-energy, which go beyond the Hartree-Fock approximation of the nucleon self-energy used to derive $g^{(HF)}$. The momentum distribution $n_{lj}$ can then be calculated easily from the imaginary part of the single-particle Green's function by

$$n(k) = \sum_{lj} 2(2j+1) \int_{-\infty}^{\epsilon_F} d\omega\, S_{lj}(k,\omega) = \sum_{lj} 2(2j+1) \int_{-\infty}^{\epsilon_F} d\omega\, \frac{1}{\pi}\mathrm{Imag}[g_{lj}(k,k;\omega)]. \tag{6}$$

The details on the evaluation of the Hartree-Fock approximation to the Green's function $g_{lj}^{(HF)}(k_1, k_2; \omega)$, the definition of $\Delta\Sigma_{lj}$ and the technique used to solve the Dyson eq.(5) will be presented in subsections II A and II B below.

### A. Model Space and Effective Hamiltonian

As outlined in the introduction, the main point of this investigation is to study effects of long range correlations by means of the Green's function approach within a finite model space. This model space shall be defined in terms of shell-model configurations including oscillator single-particle states up to the sdg shell. The oscillator parameter, $b = 1.76$ fm, has been chosen appropriate for the nucleus $^{16}$O. This model space does not allow the description of short-range correlations. Nevertheless, we also have to take into account the effects of short-range correlations by introducing an effective interaction, i.e. a $G$-matrix appropriate for the model space. This truncation of the Hilbert space into a model space, the degrees of which are treated explicitly, and the space outside this model space, which is taken into account by means of effective operators, is often referred to as a double partitioned Hilbert space and has been used before for finite nuclei [15,16] and nuclear matter [17,18].

The $G$-matrix is determined as the solution of the Bethe-Goldstone equation

$$\mathcal{G} = V + V \frac{Q_{\mathrm{mod}}}{\omega - Q_{\mathrm{mod}} T Q_{\mathrm{mod}}} \mathcal{G}, \tag{7}$$

where $T$ is identified with the kinetic energy operator, while $V$ stands for the bare two-body interaction. For the latter we have chosen the Reid soft-core potential [19]. In this equation the Pauli operator $Q_{mod}$ is defined in terms of our harmonic oscillator single-particle states. Thus applying $Q_{\text{mod}}$ to two-particle states $|\alpha\beta>$ one obtains

$$Q_{\text{mod}}|\alpha\beta> = \begin{cases} 0 & \text{if } \alpha \text{ or } \beta \text{ below Fermi level} \\ 0 & \text{if } \alpha \text{ and } \beta \text{ in model space} \\ |\alpha\beta> & \text{else} \end{cases} \quad (8)$$

The model space used in the eq. (7 and 8) includes all states up to the sdg-shell. Note that with this definition of $Q_{\text{mod}}$ we ensure that no doublecounting of correlations occurs between the $\mathcal{G}$-matrix and the evaluation of the Green's function.

In the solution of the Bethe-Goldstone eq.(7) we have chosen a constant value of $\omega = -30$ MeV for the starting energy. This value is a reasonable mean value for the sum of two single-particle energies for hole states in $^{16}$O. Clearly, this choice for a constant starting energy is an approximation introduced to simplify the calculations, but one has to note that our results do not depend significantly on the actual value of $\omega$. The use of a constant starting energy also implies that we do not try to account for a depletion of the occupation probability due to scattering into states outside the model space as it has been done e.g. in [20,21].

The matrix elements of $\mathcal{G}$ are represented in a basis of plane-wave states and are given in the form

$$< k'l'SJ_SKLT|\mathcal{G}|k''l''SJ_SKLT> \quad (9)$$

where $k', k''$ denote the relative momenta, $l', l''$ the orbital angular momentum for the relative motion, $K$ and $L$ the corresponding quantum numbers for the center of mass motion, $S$ and $T$ stand for the total spin and isospin of the interacting pair of nucleons and by definition the angular momentum $J_S$ is obtained from coupling the orbital angular momentum of relative motion and the spin $S$. The calculation of matrix elements for $\mathcal{G}$, which is appropriate for a specific finite nucleus and a corresponding model space, in this plane-wave basis has been made possible by extending the techniques for solving the Bethe-Goldstone equation for finite systems as described in [22].

### B. Nucleon Self-Energy and Green's Function

The calculation of the self-energy is performed in terms of two-particle states, characterized by single-particle momenta in the laboratory frame. Such a antisymmetrized 2-particle state would be described by quantum numbers such as

$$|k_1 l_1 j_1 k_2 l_2 j_2 JT> \quad (10)$$

where $k_i$, $l_i$ and $j_i$ refer to momentum and angular momenta of particle $i$ whereas $J$ and $T$ define the total angular momentum and isospin of the two-particle state. The transformation from the relative and c.m. coordinates, in which the matrix elements of $\mathcal{G}$ are defined (see eq.(9)) to the states displayed in eq.(10) can be made by use of the well known vector bracket transformation coefficients [23,24].

Performing an integration over one of the $k_i$, one obtains a 2-particle state in a mixed representation of one particle in a bound harmonic-oscillator state while the other is in a plane-wave state

$$|n_1 l_1 j_1 k_2 l_2 j_2 JT> = \int_0^\infty dk_1 \; k_1^2 \; R_{n_1 l_1}(b\,k_1) \; |k_1 l_1 j_1 k_2 l_2 j_2 JT> . \quad (11)$$

Here $R_{n_1 l_1}$ stands for the radial oscillator function and the oscillator length $b = 1.76$ fm has been selected which corresponds to an oscillator energy of $\hbar\omega = 13.3$ MeV. Therefore the oscillator functions are quite appropriate to describe wave functions of the bound single-particle states in $^{16}$O. Now with the help of eqs.(9 - 11) we can write down our Hartree-Fock approximation for the self-energy in momentum representation

$$\Sigma^{HF}_{l_1 j_1}(k_1, k_1') = \frac{1}{2(2j_1+1)} \sum_{n_2 l_2 j_2 JT} (2J+1)(2T+1) \, \langle k_1 l_1 j_1 n_2 l_2 j_2 JT | \mathcal{G} | k_1' l_1 j_1 n_2 l_2 j_2 JT \rangle . \quad (12)$$

The summation over the oscillator quantum numbers is restricted to the states occupied in the IPM of $^{16}$O. Clearly, this Hartree-Fock part of the self-energy is real and does not depend on energy. One obtains the HF single-particle wave functions by expanding them

$$|\alpha^{HF} ljm> = \sum_i |K_i ljm><K_i|\alpha^{HF}>_{lj} \tag{13}$$

in a complete and orthonormal set of regular basis functions within a spherical box of radius $R_{box}$ which is large as compared to the radius of the nucleus

$$\Phi_{iljm}(\mathbf{r}) = \langle \mathbf{r}|K_i ljm\rangle = N_{il} j_l(K_i r) \mathcal{Y}_{ljm}(\vartheta\varphi) \tag{14}$$

where $\mathcal{Y}_{ljm}$ denotes the spherical harmonics including the spin degrees of freedom while $j_l$ stands for the spherical Bessel functions with discrete momenta $K_i$ determined from the boundary condition

$$j_l(K_i R_{\text{box}}) = 0. \tag{15}$$

Using the normalization constants

$$N_{il} = \begin{cases} \frac{\sqrt{2}}{\sqrt{R_{\text{box}}^3} j_{l-1}(K_i R_{\text{box}})}, & \text{for } l > 0 \\ \frac{i\pi\sqrt{2}}{\sqrt{R_{\text{box}}^3}}, & \text{for } l = 0, \end{cases} \tag{16}$$

the basis function defined in eq.(14) are orthogonal and normalized within the box. The expansion coefficients of eq.(13) are obtained by diagonalizing the HF Hamiltonian

$$\sum_{n=1}^N H_{in}^{0,lj} \langle K_n|\alpha^{HF}\rangle_{lj} = \epsilon_{\alpha lj}^{HF} \langle K_i|\alpha^{HF}\rangle_{lj}. \tag{17}$$

with the matrix elements of the HF hamiltonian

$$H_{in}^{0,lj} = \langle K_i|\frac{K_i^2}{2m}\delta_{in} + \Sigma_{lj}^{HF}|K_n\rangle. \tag{18}$$

Here and in the following the set of basis states in the box has been truncated by assuming an appropriate $N$. From the HF wave functions and energies one can construct the HF approximation to the single-particle Green's function in the box, which comes out as

$$g_{\alpha lj}^{(HF)}(k,k';\omega) = \frac{<k|\alpha^{HF}>_{lj}<\alpha^{HF}|k'>_{lj}}{\omega - \epsilon_{\alpha lj}^{HF} \pm i\eta}. \tag{19}$$

Note that by choosing especially this basis we are able to separate contributions from different momenta to the HF single-particle state, which is essential in order to compute at the end of our formalism the single-particle Green's function in momentum space.

The next step in the evaluation of the irreducible self-energy is to take into account terms of second order in the effective interaction which correspond to intermediate 2-particle 1-hole (2p1h) states as displayed in Fig.1a)

$$\Sigma_{lj}^{(2p1h)}(k,k',\omega) = \frac{1}{2} \sum_{h<F} \sum_{p_1,p_2>F} \frac{<kh|\mathcal{G}|p_1 p_2><p_1 p_2|\mathcal{G}|k'h>}{\omega - e(p_1,p_2,h) + i\eta} \tag{20}$$

and intermediate 2-hole 1-particle (2h1p) states

$$\Sigma_{lj}^{(2h1p)}(k,k',\omega) = \frac{1}{2} \sum_{p>F} \sum_{h_1,h_2<F} \frac{<kp|\mathcal{G}|h_1 h_2><h_1 h_2|\mathcal{G}|k'p>}{\omega - e(h_1,h_2,p) - i\eta}. \tag{21}$$

Here we have introduced the abbreviation

$$e(\alpha,\beta,\gamma) = \epsilon_\alpha^{HF} + \epsilon_\beta^{HF} - \epsilon_\gamma^{HF} \tag{22}$$

where the $\epsilon_{\alpha,\beta,\gamma}^{HF}$ are the HF single-particle energies. Note that the summations in eq.( 20, 21) on particle labels like $p_1, p_2$ and $p$ are restricted to those single-particle states within the model space, which are above the Fermi level ($F$), whereas the labels $h_1, h_2$ and $h$ refer to hole states.

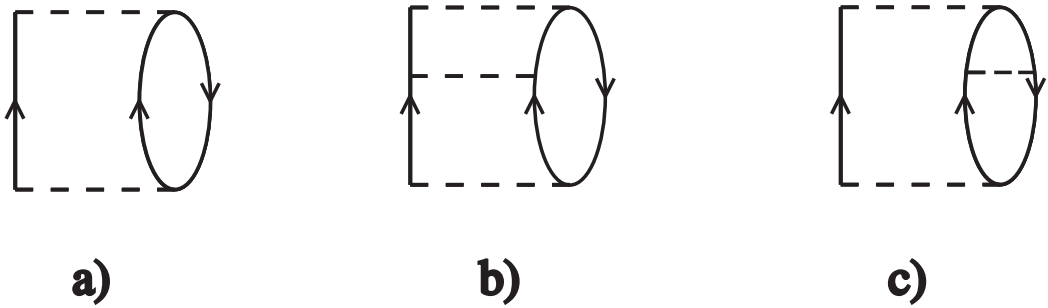

FIG. 1. Contributions to the nucleon self-energy of second (a) and third order in the interaction (b) and (c)

After the definition of the self-energy we can now proceed and calculate the corresponding single-particle Green's function $g_{lj}$ by solving a Dyson equation ( see eq.(5)) with the $g^{HF}$ taken from eq.(19) and including the correlation effects contained in $\Delta\Sigma_{lj}$

$$\Delta\Sigma_{lj}(k,k',\omega) = \Sigma_{lj}^{(2p1h)}(k,k',\omega) + \Sigma_{lj}^{(2h1p)}(k,k',\omega) . \tag{23}$$

### C. Solution of the Dyson Equation

The technique we use to solve the Dyson equation in order to extract the basic ingredients of the single-particle Green's function is very similar to the one developed in [16,25]. So we will restrict ourselves in giving a short review of the basic steps towards the determination of the single-particle Green's function for a finite system within a model space of discrete single-particle states.

In order to obtain the information necessary for the Lehmann representation of the Green's function, we rewrite the Dyson equation as an eigenvalue problem [16]

$$\begin{pmatrix} H_{11}^0 & \ldots & H_{1N}^0 & a_{11} & \ldots & a_{1P} & A_{11} & \ldots & A_{1Q} \\ \vdots & & \vdots & \vdots & & \vdots & \vdots & & \vdots \\ H_{N1}^0 & \ldots & H_{NN}^0 & a_{N1} & \ldots & a_{NP} & A_{N1} & \ldots & A_{NQ} \\ a_{11} & \ldots & a_{N1} & e_1 & & 0 & & & \\ \vdots & & \vdots & & \ddots & & & & \\ a_{1P} & \ldots & a_{NP} & 0 & & e_P & & & 0 \\ A_{11} & \ldots & A_{N1} & & & & E_1 & & \\ \vdots & & \vdots & & & & & \ddots & \\ A_{1Q} & \ldots & A_{NQ} & 0 & \ldots & 0 & & \ldots & E_Q \end{pmatrix} \begin{pmatrix} X_{0,k_1}^n \\ \vdots \\ X_{0,k_N}^n \\ X_1^n \\ \vdots \\ X_P^n \\ Y_1^n \\ \vdots \\ Y_Q^n \end{pmatrix} = \omega_n \begin{pmatrix} X_{0,k_1}^n \\ \vdots \\ X_{0,k_N}^n \\ X_1^n \\ \vdots \\ X_P^n \\ Y_1^n \\ \vdots \\ Y_Q^n \end{pmatrix} , \tag{24}$$

where for simplicity we have dropped the corresponding conserved quantum numbers for parity and angular momentum $(lj)$. The matrix to be diagonalized contains the HF hamiltonian $(H_{ij}^0)$ defined in (18) and the coupling to the $P$ different $2p1h$ configurations and $Q$ $2h1p$ states which can be constructed in our model space with quantum numbers for parity and angular momentum $j$, which are compatible to the single-particle quantum numbers $lj$ under consideration. As long as we are still ignoring any residual interaction between the various $2p1h$ and $2h1p$ configurations the corresponding parts of the matrix in (24) is diagonal with diagonal elements defined by $e_i$ ($E_j$) for $2p1h$ ($2h1p$)

$$e_i = e(p_1, p_2, h) \qquad E_j = e(h_1, h_2, p) , \tag{25}$$

where we have used again the abbreviation (22). The matrix elements connecting the HF part to the additional states refer to

$$\begin{aligned} a_{mi} &= < k_m h | \mathcal{G} | p_1 p_2 > \\ A_{mj} &= < k_m p | \mathcal{G} | h_1 h_2 > \end{aligned} \tag{26}$$

Solving the eigenvalue problem (eq.(24)) one gets as a result the single-particle Green's function in the Lehmann representation in the discrete basis of the box defined in eq.(14). The eigenvalues $\omega_n$ define the position of the poles of the Green's function, which refer to the various states of the system with $A \pm 1$ nucleons and the corresponding spectroscopic amplitudes are given by

$$< \Psi_0^A |a_{k_i}| \Psi_n^{A+1} > = X_{0,k_i}^n \quad for \quad \omega_n > E_F$$
$$< \Psi_0^A |a_{k_i}^\dagger| \Psi_n^{A-1} > = X_{0,k_i}^n \quad for \quad \omega_n < E_F \quad (27)$$

which depend on whether $\omega_n$ is an energy above or below the Fermi energy $E_F$. With the help of this nomenclature we can finally write down the momentum distribution for a given partial wave

$$n_{lj}(k_i) = \sum_n \Theta(E_F - \omega_n) \, |X_{0,k_i}^n|^2 \, . \quad (28)$$

In a straightforward way one can improve the approximation discussed so far and incorporate the effects of residual interactions between the $2p1h$ configurations as illustrated in the diagrams displaying the self energy in Fig. 1b) and c). The same holds for the $2h1p$ configurations. One simply has to modify the corresponding parts of the matrix in eq.(24) and replace

$$\begin{pmatrix} e_1 & \cdots & 0 \\ \vdots & \ddots & \\ 0 & \cdots & e_P \end{pmatrix} \Longrightarrow \mathcal{H}_{2p1h} \, , \quad \text{and} \quad \begin{pmatrix} E_1 & \cdots & 0 \\ \vdots & \ddots & \\ 0 & \cdots & E_Q \end{pmatrix} \Longrightarrow \mathcal{H}_{2h1p} \, , \quad (29)$$

where $\mathcal{H}_{2p1h}$ and $\mathcal{H}_{2h1p}$ contain the residual interactions in the $2p1h$ and $2h1p$ subspaces. The solution of the eigenvalue problem also leads to a normalization condition, which ensures that

$$\sum_n |X_{0,k_i}^n|^2 + \sum_m |X_{0,k_i}^m|^2 = \sum_n | < \Psi_n^{A+1}|a_{k_i}^\dagger|\Psi_0^A > |^2 + \sum_m | < \Psi_m^{A-1}|a_{k_i}|\Psi_0^A > |^2 = 1 \, , \quad (30)$$

where the sum on $n$ accounts for all solutions with $\omega_n$ larger than the Fermi energy and the sum on $m$ for all solutions with $\omega_n$ below the Fermi energy. Again this implies that one ignores all effects of correlations, which are due to configurations outside the model space, like e.g. an effective energy-dependent hamiltonian [20,21]. In that case one has to renormalize the condition of eq.(30) as well.

Note that for the solution of the eigenvalue problem one can apply the so-called "BAsis GEnerated by Lanczos" (BAGEL) scheme [16,25,26] in order to get a very efficient representation of the single-particle Green's function in terms of a few "characteristic" poles in the Lehmann representation.

### D. Ground-State Properties

The single-particle Green's function calculated by the techniques described above, may also be used to determine expectation values of arbitrary single-particle operators

$$< \Psi_0 |\hat{O}| \Psi_0 > = \int_C \frac{d\omega}{2\pi i} \sum_{\alpha\beta} < \alpha|\hat{O}|\beta > G_{\alpha\beta}(\omega) \, . \quad (31)$$

The $C$ below the integral sign denotes a contour integration counter-clockwise in the upper half plane including the real axis. Therefore only the contributions from the poles at energies below the Fermi energy have to be considered and using the nomenclature of the matrix representation in eq.(24) one can rewrite this expectation value, now in the case of the square radius, as

$$< \Psi_0^A |r^2| \Psi_0^A > = \sum_{lj} 2(2j+1) \sum_{i,n=1}^{Nmax} \sum_{m<F} < k_i|r^2|k_n >_l X_{0,k_i}^m X_{0,k_n}^m \, , \quad (32)$$

where the sum on $m$ is restricted to solutions with energies $\omega_m$ below the Fermi energy and the factor of 2 accounts for isospin degeneracy. The matrix elements for $r^2$ are given by

$$< k_i|r^2|k_n >_l = N_{il} N_{nl} \int_0^{Rbox} dr \, r^4 \, j_l(k_i r) \, j_l(k_n r) \, . \quad (33)$$

The total energy of the ground state is obtained from the "Koltum sum rule" which reads in our case as

$$E_0^A = \sum_{lj} 2(2j+1) \sum_{i=1}^{Nmax} \sum_{m<F} \frac{1}{2} \left(\frac{k_i^2}{2m} + \omega_m\right) S(k_i, \omega_m) \quad (34)$$

and again $\omega_m$ is restricted to energies below the Fermi energy.

## III. RESULTS AND DISCUSSION

The spectral function $S_{lj}(k,\omega)$ contains the information on the energy and momentum distribution of the single-particle strength in the various partial waves $lj$. As a first result we would like to discuss the distribution of the single-particle strength with respect to the energy as derived from our investigation of long-range correlations. A useful measure for this energy distribution in the various partial waves is defined by

$$\mathcal{N}_{lj}(E^*) = \int_{\epsilon_F - E^*}^{\epsilon_F} d\omega \int_0^\infty dk\ S_{lj}(k,\omega). \quad (35)$$

This energy-distribution function represents the single-particle strength for a given partial wave $lj$, which should be observed if reactions to all states in the residual (A-1) particle system are included up to an excitation energy of $E^*$ above the corresponding ground state. These distribution functions cannot directly be identified with occupation probabilities. Occupation probabilities refer to one specific single-particle state like e.g. the $0p_{1/2}$ oscillator state and therefore they must always be less equal one for a system of fermions. The energy-distribution function defined above refers to all states for a given set of $lj$ quantum numbers, not distinguishing the various radial dependencies in configuration or momentum space.

Results for these energy-distribution functions for the nucleus $^{16}O$ are shown in Table I. The spectral strength in the partial waves with $l = 1$ is dominated in each case by the transition to one state in the residual nucleus, the quasi-particle state. In the case of the $p_{1/2}$ state this is the ground state of $^{15}N$ (or $^{15}O$) and for the $p_{3/2}$ partial wave this is the first excited state with $j = 3/2$. These quasi-particle states represent a spectral strength of 0.835 and 0.853 for $p_{1/2}$ and $p_{3/2}$, respectively. The remaining single-particle strength of 0.085 in the case of $p_{1/2}$ (0.088 for $P_{3/2}$) is distributed over many states in a broad window of excitation energies ranging up to $E^*$ around 100 MeV.

TABLE I. The energy-distribution function $\mathcal{N}_{lj}$, defined in eq.(35) for various excitation energies $E^*$ in different partial waves. The values marked by $^+$ in the $l = 1$ partial waves denote the spectroscopic strength for the corresponding quasi-hole state

| $lj$ $E^*$ | with residual interaction | | | without residual interaction | | |
|---|---|---|---|---|---|---|
| | 20 MeV | 40 MeV | total | 20 MeV | 40 MeV | total |
| $s_{1/2}$ | 0.265 | 0.756 | 0.993 | 0.368 | 0.778 | 0.987 |
| $p_{3/2}$ | 0.853$^+$ | 0.910 | 0.941 | 0.887$^+$ | 0.917 | 0.953 |
| $p_{1/2}$ | 0.835$^+$ | 0.871 | 0.920 | 0.869$^+$ | 0.891 | 0.939 |
| $d_{5/2}$ | 0.020 | 0.028 | 0.039 | 0.005 | 0.019 | 0.029 |

TABLE II. Results for binding energy per nucleon (E/A), radius of the nucleon distribution ($<r>$) and particle number $A$ of $^{16}O$ for the HF approximation, the Green's function approach including diagrams of second order (compare Fig.1a) in the self-energy, and with inclusion of residual interaction (see e.g. Fig.1b and c). Since the calculations do not include the Coulomb interaction between protons, the effects of the Coulomb interaction have also been removed from the experimental data.

| | E/A [MeV] | $<r>$ [fm] | A |
|---|---|---|---|
| Hartree-Fock | -3.632 | 2.522 | 16.000 |
| plus second order | -6.146 | 2.766 | 15.973 |
| with residual interaction | -6.265 | 2.817 | 16.028 |
| Experiment | -9.118 | 2.580 | 16.000 |

The single-particle strength is much less located in the case of the $s_{1/2}$ partial wave. This can be seen from Fig. 2 which shows the contributions to the energy-distribution function for this partial wave, accumulated in energy bins of 5 MeV. Non-negligible contributions are obtained up to excitation energies of 75 MeV, smaller contributions are located at even slightly higher energies. Also the spectral strength for the $d_{5/2}$ partial wave, which we have included in Table I as one example of a partial wave which in the IPM does not carry any strength at all, is distributed over many states at various energies.

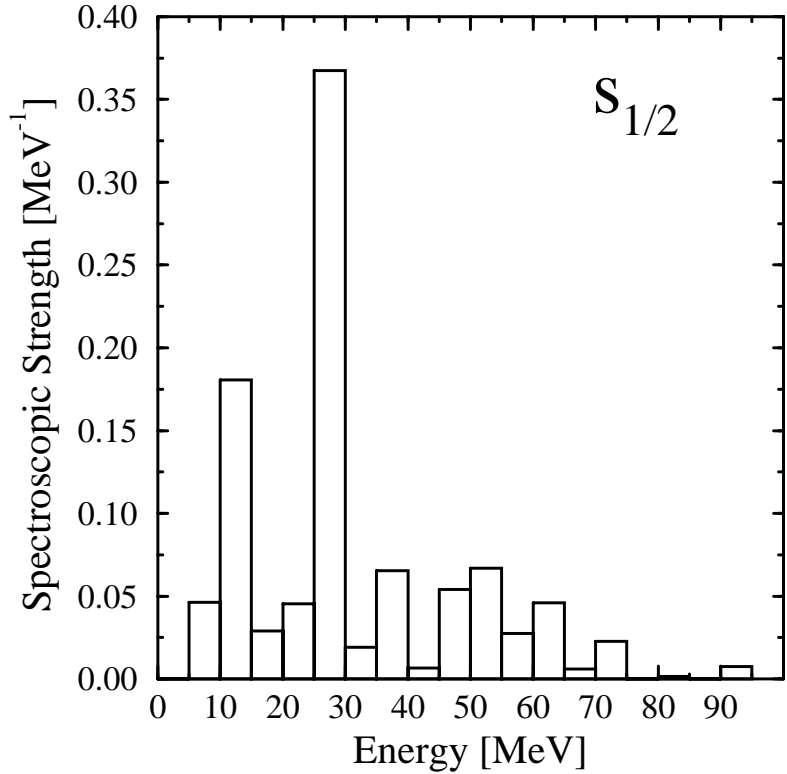

FIG. 2. Spectral strength in the partial wave $s_{1/2}$ as a function of the excitation energy $E^*$. The contributions to the energy-distribution function defined in eq.(35) are collected in energy bins of 5 MeV.

It has been one of our aims to explore the importance of contributions to the self-energy represented by diagrams of third and higher order in the residual interaction as represented by Fig. 1b) and c). As it has been discussed above, these contributions are easily taken into account by including the residual interaction between the $2p1h$ and $2h1p$ configurations in the corresponding parts of the matrix in eq.(24). The results shown in the left part of Table I which we discussed so far, were obtained including this residual interaction. On the other hand, those results shown in the right part of the table are deduced from calculations in which these residual interactions are ignored. This means that they approximate the self-energy by terms of first and second order in the interaction $G$.

The energy-distribution functions show only a small sensitivity with respect to the residual interaction. One can observe a small additional depletion of the total strength in $p$ states accompanied by a slight enhancement of the strength in the $d$-states. This means that the residual interaction terms slightly enhance the deviation from the IPM model which is observed if only terms up to second order are taken in the definition of the self-energy. Also one observes a small removal of strength from the quasi-particle states in the $p$ states to other states at small excitation energies.

The next question, which we would like to address is the momentum distribution of the spectral function at various energies. As a first example we present in Fig.3 the momentum distribution for the $p_{1/2}$ partial wave. In order to visualize the energy dependence we plotted the integrated strength for different excitation energy limits $E^*$

$$n_{lj}(k) = \int_{\epsilon_F - E^*}^{\epsilon_F} d\omega \, S_{lj}(k,\omega) \tag{36}$$

The dashed line of this figure refers to the momentum distribution for the quasi-particle or quasi-hole state, which is the transition to the ground state of the nucleus $^{15}$N. The dashed-dotted and solid line are obtained if the spectral strength is included up to an excitation energy $E^*$ of 40 MeV and 80 MeV, respectively. The solid line refering to $E^*$ of 80 MeV represents the total momentum distribution.

In order to compare our data we also plotted the Hartree-Fock result (see eq.(19)), hence our mean field description (dotted line). This figure displays quite clearly that the high-momentum components in the spectral function for the quasi hole state, which is observed in a nucleon knockout reaction to the ground state of the remaining nucleus $^{15}$N are very well described already within the Hartree-Fock approach. The inclusion of low-energy correlations has hardly any effect on these high-momentum components.

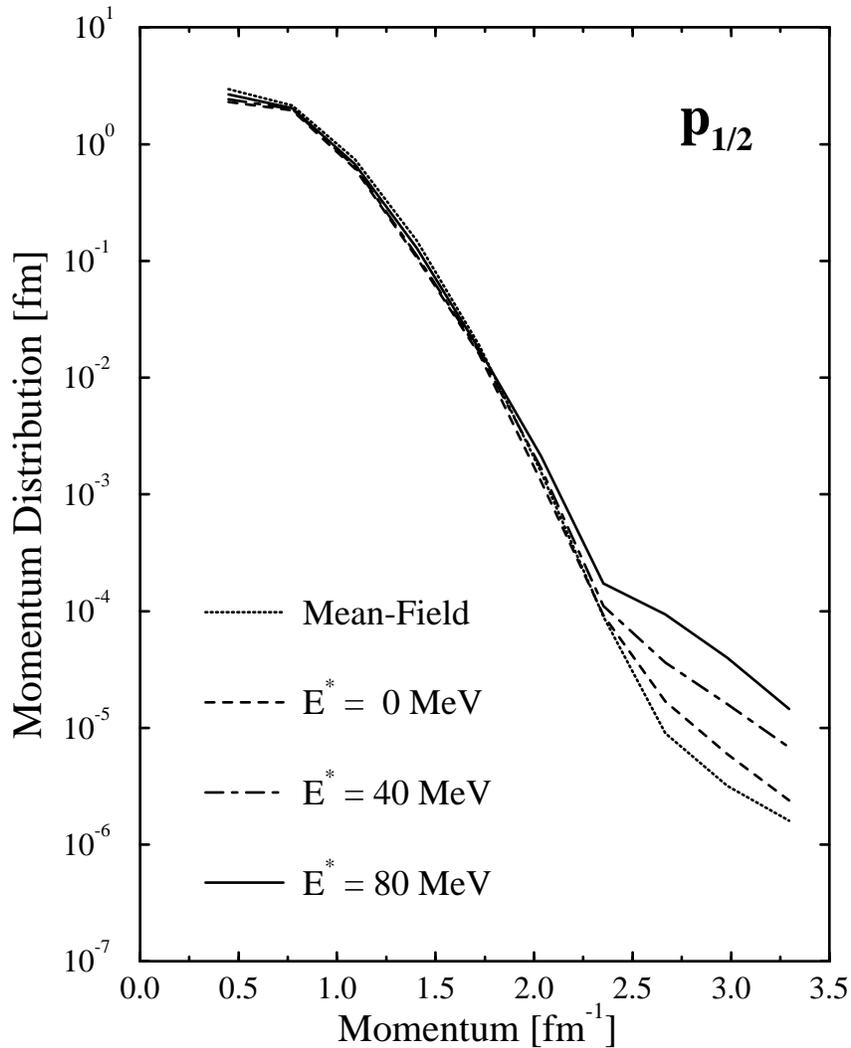

FIG. 3. Momentum distribution in the $p_{1/2}$ partial wave evaluated with various limits in the excitation energy $E^*$, as defined in eq.(36).

The long-range correlations described in terms of the excitation modes lead to an increase of the total momentum distribution at high momenta. This increase, however, is observed only if the spectral strength up to excitation energies of 40 or 80 MeV is included. The same features are observed also for the other partial waves included in our analysis. To visualize in more detail this collection of high-momentum components in the momentum distribution as a function of the maximal excitation energy $E^*$ included in the energy integral of eq.(36), we show in Fig. 4 the momentum distribution integrated over all momenta $k > 2.4$ fm$^{-1}$ as a function of the excitation energy $E^*$. Since we are particularly interested in the importance of the residual interaction between the excitation modes of the model space (see also discussion of Table I above) we compare in this figure the results obtained with and without inclusion of this residual interaction. As an example Fig. 4 shows the momentum distribution for the $p_{3/2}$ partial wave. We can see very clearly that the residual interaction yields a slight shift of the high-momentum components to lower energies. This effect, however, is not very large, supporting the discussion of the energy-distribution function above.

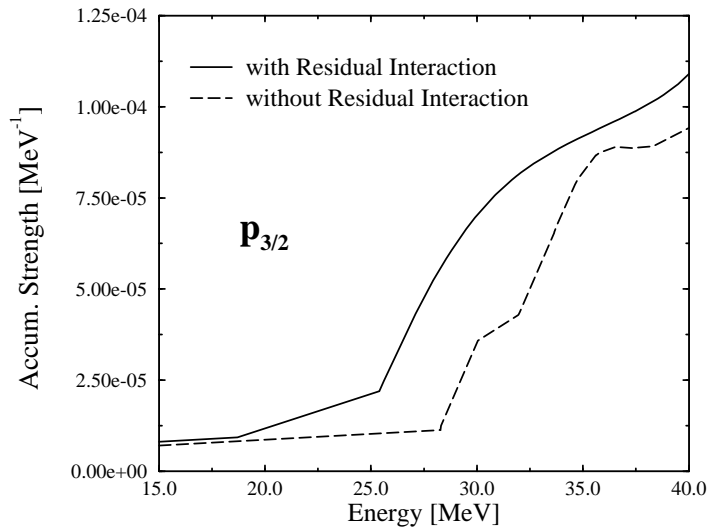

FIG. 4. Integral of the momentum distribution for momenta $k > 2.4$ fm$^{-1}$ in the $p_{3/2}$ partial wave as a function of the excitation energy $E^*$ limiting the integral in eq.(36).

Now, comparing our results with the experimental data obtained in a recent $(e,e'p)$ experiment on $^{16}$O done at MAMI (Mainz) [1], we are in good agreement in the sense that the inclusion of long-range correlations does not lead to any significant modification of the momentum distribution for the quasi-hole state, i.e. the spectral function leading to the ground state. Therefore the momentum distribution should be well described by the IPM or Hartree-Fock approximation, as it has been observed in the analysis of those data.

For the heavy nucleus $^{208}$Pb, however, a spectral function has been deduced from the experiment [2], which is larger for high momenta than the IPM prediction for the quasi-hole state. What is the difference between these nuclei? Since the numerical effort prevents a similar calculation for the nucleus $^{208}$Pb, we try to simulate the higher density of single-particle states observed in $^{208}$Pb as compared to $^{16}$O by reducing the spacing between the single-particle energies obtained in the HF approximation by a common quenching factor. This quenching reduces the energies of the $2p1h$ and $2h1p$ configurations. This is of course a very crude approximation to simulate a calculation on $^{208}$Pb. It ignores the fact that the single-particle states around the Fermi energy for $^{208}$Pb have quite a different structure from those in $^{16}$O, which implies that also the matrix elements of the two-body interaction will be different. Nevertheless, the results of this simple estimate can give us a hint on the possible effect of long-range correlations on the momentum distribution for heavier systems.

Fig. 5 displays the high-momentum component of the spectral strength for the quasi-hole state in the $p_{3/2}$ channel as a function of the quenching factor used to modify the single-particle spectrum. The spectral strength of the high-momentum components has been defined by the spectral function integrated over all momenta $k > 2.4$ fm$^{-1}$. We observe a systematic increase of this high-momentum part, if the spacing between the single-particle energies is reduced. This is an indication that the larger collectivity of the low-energy states in $^{208}$Pb could be responsible for a significant enhancement of the momentum distribution for the spectral function of the quasi-hole state at high momenta.

Finally, we would like to discuss the effects of the $2p1h$- and $2h1p$- configurations and the residual interaction in between on the ground-state properties of $^{16}$O. For this purpose we list in Table II the result for the binding energy per nucleon, calculated according to eq.(34) divided by the particle number $A$ and the root mean square radius for the distribution of the nucleons $<r>$. The table shows results obtained in the Hartree-Fock (HF) approximation and in the present Green's function approach with and without inclusion of the residual interaction between $2p1h$ and $2h1p$ configurations. The binding energy obtained in the HF approximation is rather small as compared to the experimental value. One should recall that our HF approximation actually corresponds to a calculation using a $G$ matrix evaluated with a Pauli operator as defined in eq.(8). This Pauli operator not only forbids scattering into single-particle states occupied in the HF wave function for $^{16}$O, but also suppresses the intermediate 2-particle state in the model space, as we want to consider them in the next step. Therefore it is understandable that the HF energy is below energies typically obtained in Brueckner-Hartree-Fock (BHF) approximation.

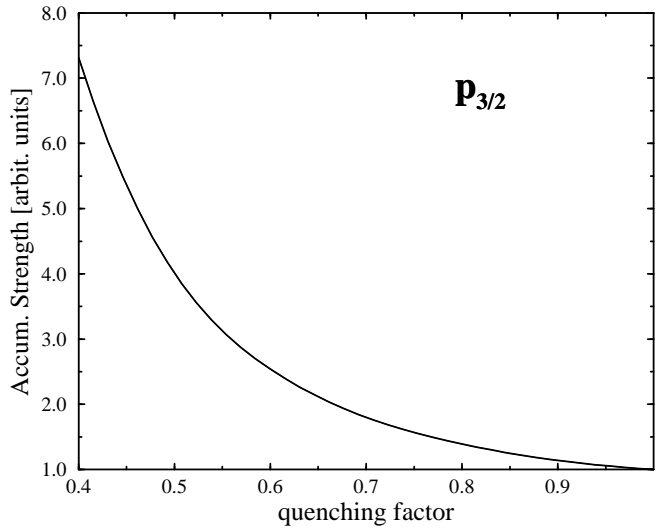

FIG. 5. Single-particle strength for the spectral function of the quasi-hole state at high momenta, $k > 2.4$ fm$^{-1}$ as a function of the factor, quenching the spacing of the single-particle energies.

The inclusion of low-lying $2p1h$ and $2h1p$ terms within our Green's function approximation increases the calculated binding energy per nucleon quite substantially. Typically such an increase of the calculated binding energy is accompanied by a decrease of the calculated radius, a phenomenon which is called the "Coester band" behavior and has originally been observed in nuclear matter [27] but is also known for finite nuclei [28]. It is interesting to note, however, that the inclusion of long-range correlations, as it is done here, yields an increase of the binding energy together with an increase of the calculated radius. Similar results have been obtained in previous calculations [25]. In those calculations, however, the results for the radii were not as reliable as they were calculated in a finite basis of oscillator states.

In the calculation of the ground-state properties we find that the contributions to the self-energy of the nucleons beyond the terms of second order yield a slight enhancement of the trends observed in the calculation, which is restricted to terms of second order in $G$. This enhancement is noticeable but not very significant. These results give rise to the hope that a microscopic nuclear structure calculation, which goes beyond the Brueckner-Hartree-Fock approach by including effects of $2h1p$ contributions and the relativistic effects of the Dirac-Brueckner-Hartree-Fock [29,30] may reproduce the empirical data from a realistic NN interaction without any free parameter.

## IV. CONCLUSIONS

The effects of long-range correlations on the momentum distribution of the nucleons in a nucleus has been studied for the example of the nucleus $^{16}$O. The spectral function, describing the momentum- and energy distribution for nucleons in the various partial waves is derived from the single-particle Green's function, which is obtained from a solution of the Dyson equation in a basis of plane-wave states. The self-energy for the nucleons entering this Dyson equation contains the Brueckner-Hartree-Fock term plus the coupling to $2p1h$ and $2h1p$ configurations. The long-range correlations are described in terms of these $2p1h$ and $2h1p$ configurations within a model space, which includes single-particle states up to the $sdg$ shell. The effect of the residual interaction between these configurations has been studied.

The inclusion of long-range correlations does not produce any significant change in the momentum distribution predicted for the quasi-hole states, i.e. for nucleon removal leading to the ground state (or lowest state for a given $lj$) of the remaining nucleus, as compared to the independent particle model (IPM). This is in agreement with the experimental data of $(e,e'p)$ experiments on $^{16}$O [1].

The collectivity of the low-lying states can be enhanced by an artificial reduction of the energy spacing between the single-particle energies. Such a quenching of the single-particle spectrum may be considered as a model for heavier nuclei with a large density of states around the Fermi energy. This yields a significant enhancement of the spectral function for the quasi-hole state at high momenta. Therefore it is plausible that long-range correlations may lead to such an enhancement for heavier nuclei, as it is observed in the experimental data of $(e,e'p)$ experiments on $^{208}$Pb [2].

The total momentum distribution, which is obtained from the spectral function by including the contributions which lead to the excited states of the (A-1) particle system as well, shows a significant enhancement at high momenta. For the $l = 1$ partial waves the spectroscopic strength for the removal of nucleons is highly concentrated in the quasi-hole state. On the other hand, a broad distribution in energy is observed for the single-particle strength with $l = 0$ and $l > 1$. The spectroscopic strength for the partial waves with $l > 1$ is identical to zero in the IPM. The effects of the residual interaction between the $2h1p$ and $2p1h$ configurations, which yields contributions to the self-energy as displayed in Fig. 1b) and c), are not very significant. This demonstrates that the nucleon self-energy is quite well represented by the terms of first and second order in the Brueckner $G$ matrix.

The inclusion of long-range correlations within the Green's function approach leads to an increase of the calculated binding energy per nucleon accompanied by an increase of the radius. This shows that a careful inclusion of $2h1p$ configurations in the nucleon self-energy provides an improvement in the microscopic calculation of ground-state properties of nuclei from realistic NN interactions.

This project has been supported by the SFB 382 (DFG, Germany) and the Graduiertenkolleg "Struktur und Wechselwirkung von Hadronen und Kernen" (DFG Mu 705/3). This support is gratefully acknowledged.

[1] K.I. Blomqvist et al., *Phys. Lett.* **344 B** (1995) 85.
[2] I. Bobeldijk et al., *Phys. Rev. Lett.* **73** (1994) 2684.
[3] S.C. Pieper, R.B. Wiringa, and V.R. Pandharipande, *Phys. Rev.* **C46** (1992) 1741.
[4] G. Co', A. Fabrocini, and S. Fantoni, *Nucl. Phys.* **A 568** (1994) 73.
[5] O. Benhar, A. Fabrocini, S. Fantoni, and I. Sick, *Nucl. Phys.* **A579** (1994) 493.
[6] H. Müther, A. Polls, and W.H. Dickhoff, *Phys. Rev.* **C 51** (1995) 3040.
[7] H. Müther, and W.H. Dickhoff, *Phys. Rev.* **C 49** (1994) R17.
[8] A. Ramos, A. Polls, and W.H. Dickhoff, *Nucl Phys.* **A503** (1989) 1.
[9] O. Benhar, A. Fabrocini, and S. Fantoni, *Nucl. Phys.* **A505** (1989) 267.
[10] B.E. Vonderfecht, W.H. Dickhoff, A. Polls, and A. Ramos, *Nucl. Phys.* **A 555** (1993) 1.
[11] H. Müther, G. Knehr, and A. Polls, *Phys. Rev.* **C 52** (1995) 2955.
[12] Z.Y. Ma and J. Wambach, *Phys. Lett.* **B 256** (1991) 1.
[13] C. Mahaux and R. Sartor, *Adv. Nucl. Phys.* **20** (1991) 1.
[14] W.H. Dickhoff and H. Müther, *Rep. Prog. Phys.* **55** (1992) 1947.
[15] B.R. Barrett and M.W. Kirson, *Adv. in Nucl. Phys.* **6** (1973) 219
[16] H. Müther and L.D. Skouras, *Nucl. Phys.* **A555** (1993) 541.
[17] T.T.S. Kuo, Z.Y. Ma and R. Vinh Mau, *Phys. Rev.* **C33** (1987) 717.
[18] M.F. Jing, T.T.S. Kuo and H. Müther, *Phys. Rev.* **C 38** (1988) 2408.
[19] R.V. Reid, *Ann. of Phys.* **50** (1968) 411.
[20] S.D. Yang and T.T.S. Kuo, *Nucl. Phys.* **A456** (1986) 413.
[21] K. Allaart, P.J. Ellis, W.J.W. Geurts, J. Hao, T.T.S. Kuo and G.A. Rijsdijk, *Phys. Rev.* **C74** (1993) 895.
[22] H. Müther and P.U. Sauer, in *Computational Nuclear Physics 2* eds. K. Langanke, J.A. Maruhn and S.E. Koonin, page 30ff (Springer Verlag N.Y. 1993).

[23] C.W. Wong and D. M. Clement, *Nucl. Phys.* **A183** (1972) 210.
[24] D. Bonatsos and H. Müther, *Nucl. Phys.* **A496** (1989) 23.
[25] H. Müther and L.D. Skouras, *Nucl. Phys.* **A581** (1995) 247.
[26] H. Müther, T. Taigel and T.T.S. Kuo, *Nucl. Phys.* **A482** (1988) 601.
[27] F. Coester, S. Cohen, B.D. Day, and C.M. Vincent *Phys. Rev.* **C1** (1970) 769.
[28] K.W. Schmid, H. Müther, and R. Machleidt, *Nucl. Phys.* **A530** (1991) 14.
[29] F. Boersma and R. Malfliet, *Phys. Rev.* **C49** (1994) 1495.
[30] R. Fritz and H. Müther, *Phys. Rev.* **C 49** (1994) 633.

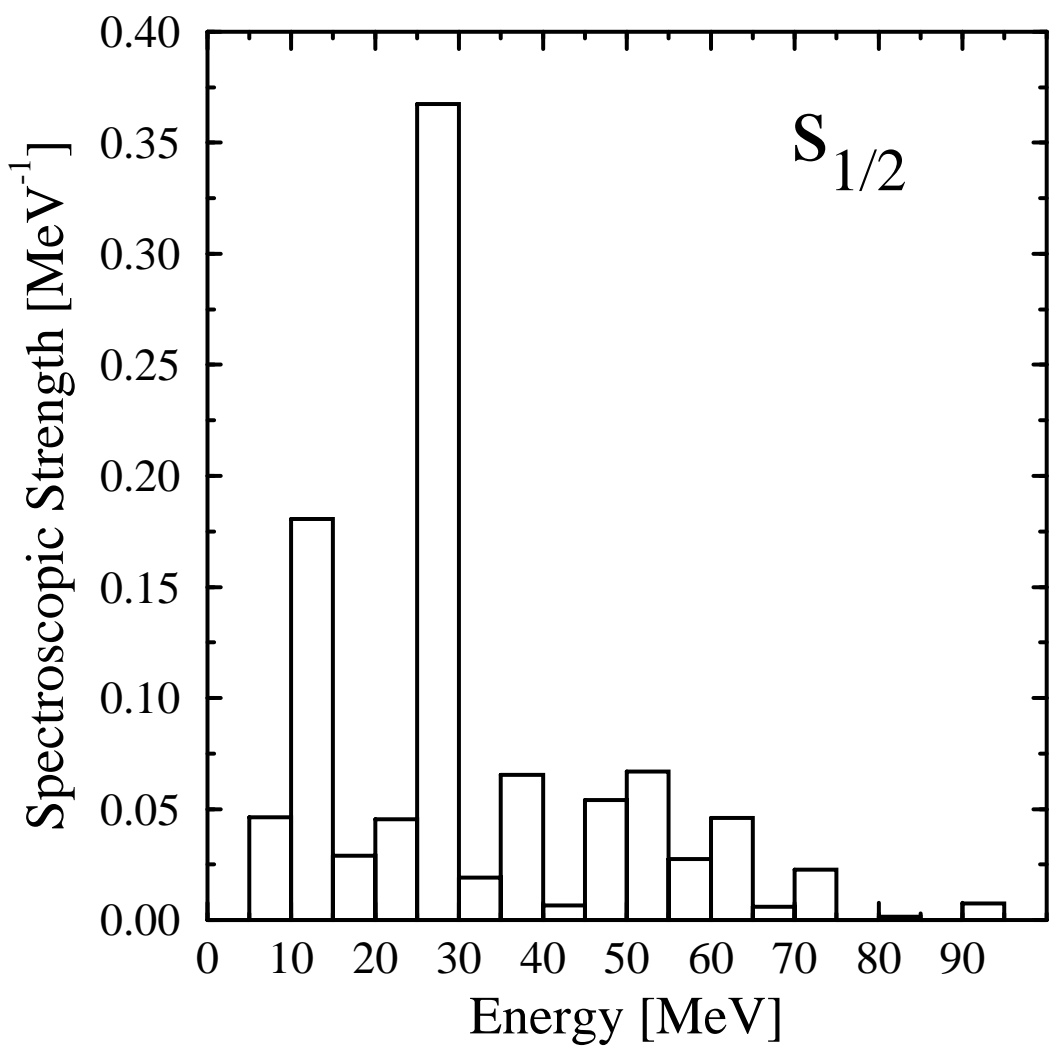

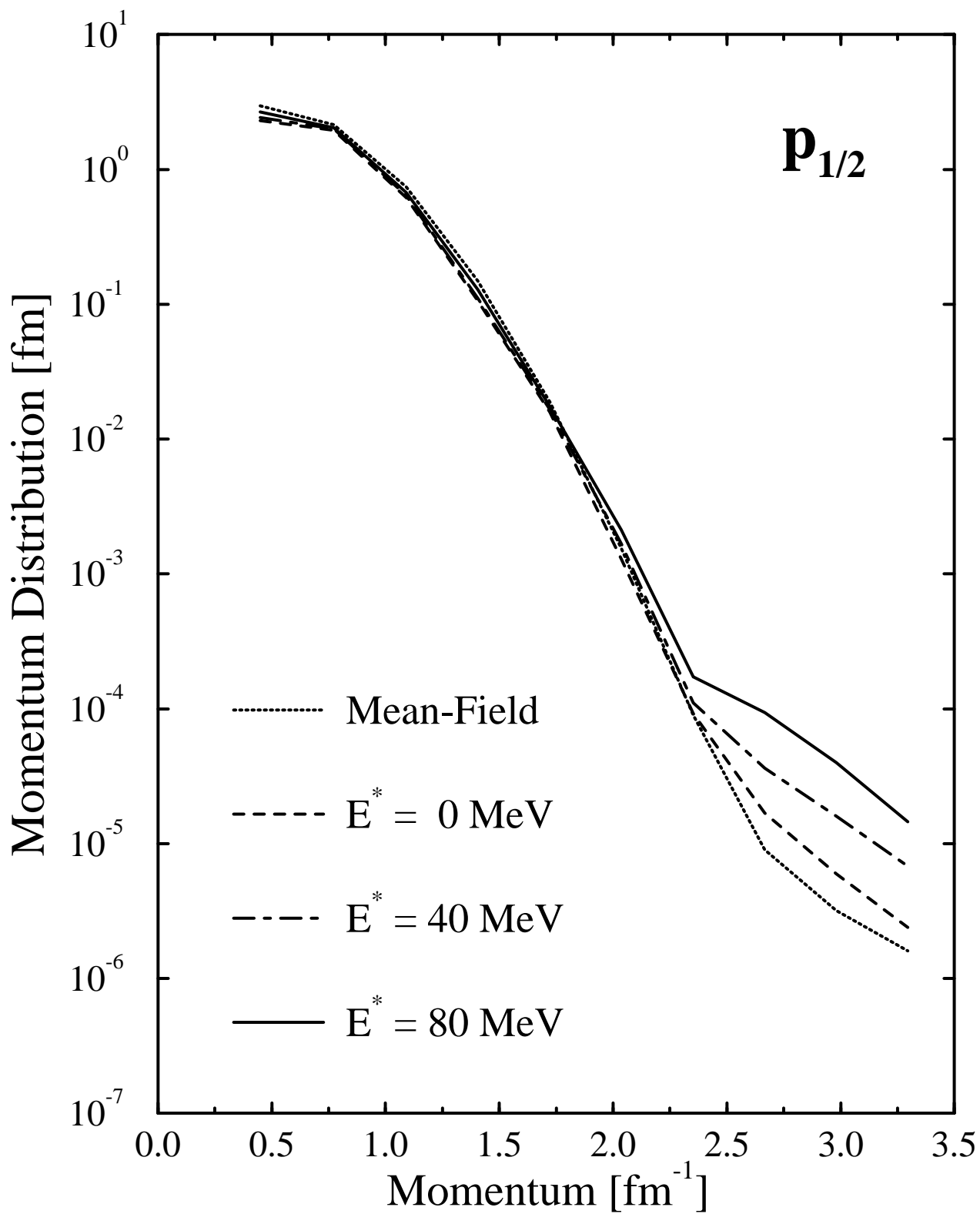

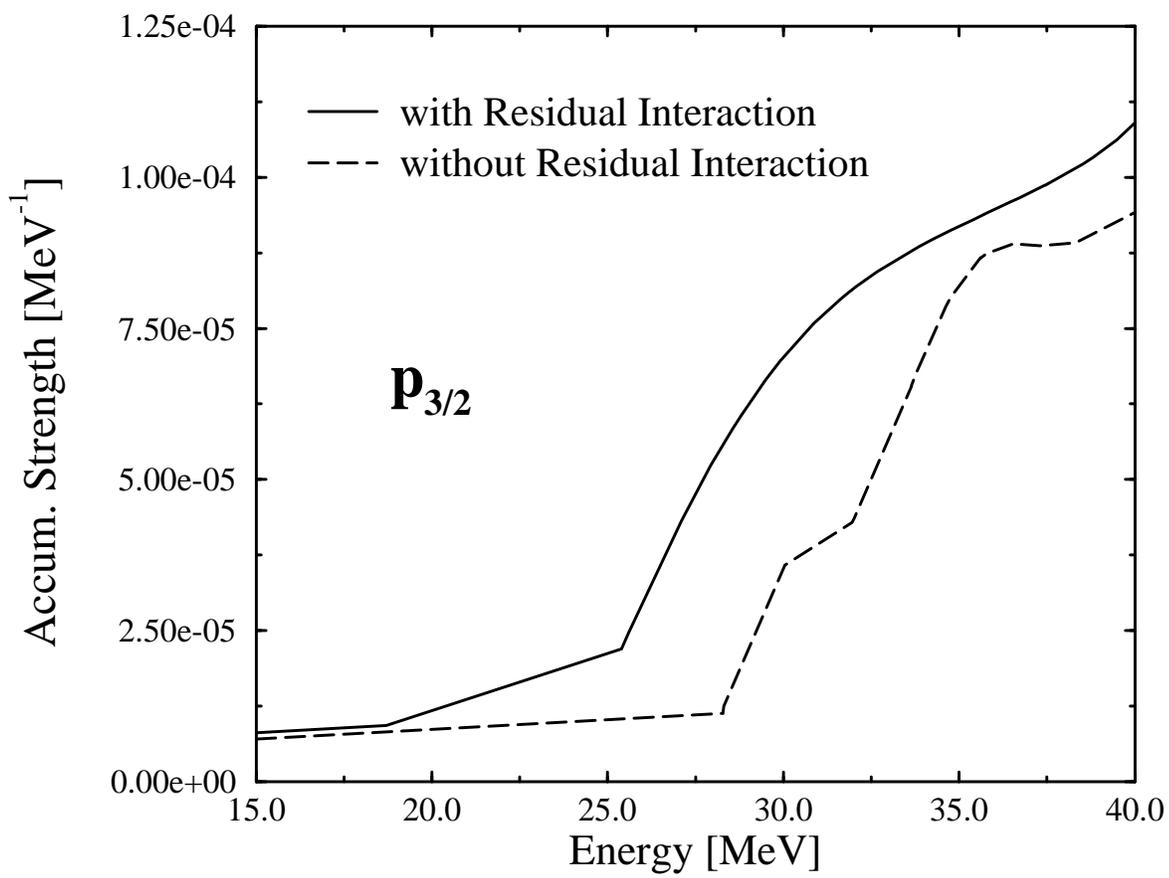

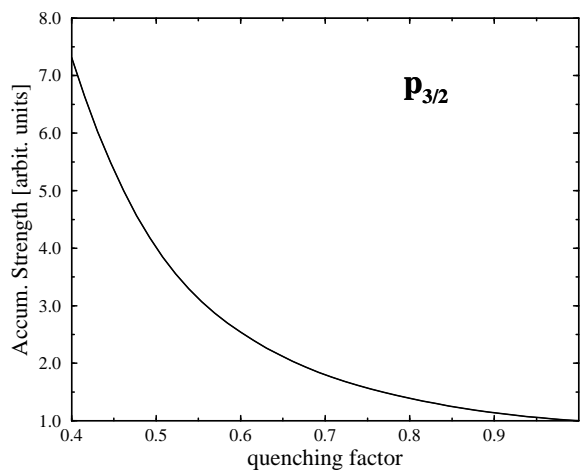